\documentclass[10pt,conference]{IEEEtran}

\usepackage[cmex10]{amsmath}
\usepackage{amsfonts}

\newtheorem{example}{Example}
 
\newtheorem{proposition}{Proposition}

\newenvironment{mproof}{\hspace{8pt}\ti{Proof:}}{~~~~QED}
\newenvironment{mmproof}{\hspace{8pt}\ti{Proof:}}{}

\newcommand{\pnt}[1]{{\mbox{\boldmath $#1$}}}
\newcommand{\cof}[2]{\mbox{$#1_{\boldsymbol{#2}}$}}

\newcommand{\s}[1]{\mbox{$\{#1\}$}}

\newcommand{\nGz}[2]{$G_{non-\{z\}}$}

\newcommand{\prr}[1]{\mi{Prev}(\boldsymbol{q})}

\newcommand{\mi}[1]{\mathit{#1}}
\newcommand{\ti}[1]{\textit{#1}}
\newcommand{\tb}[1]{\textbf{#1}}

\newcommand{\Sub}[2]{\mbox{$\mi{#1}_\mi{#2}$}}

\newcommand{\Comment}[1]{}

\newcommand{\mb}[1]{\mbox{$\mathbb{#1}$}\xspace}

\newcommand{\tr}[3]{\mbox{(\pnt{#1_{#2}},\ldots,\pnt{#1_{#3}})}}
\newcommand{\rch}[3]{\mbox{$\mi{Rch}(#1,#2,#3)$}}
\newcommand{\db}{\mbox{\textit{Ic3-db}}\xspace}
\newcommand{\Sep}{\textit{Ja-ver}\xspace}
\newcommand{\Jnt}{\textit{Jnt-ver}\xspace}
\newcommand{\Abc}{\textsc{Abc}\xspace}

\newcommand{\ja}{\textsc{Ja}\xspace}
\newcommand{\cex}{\textsc{Cex}\xspace}
\newcommand{\pdr}{\textsc{Ic3}\xspace}

\newcommand{\clauseDB}{\textit{clauseDB}\xspace} 

\usepackage[noadjust]{cite}
\usepackage{color}
\usepackage{inconsolata}
\usepackage{wrapfig}
\usepackage{graphicx}
\usepackage{enumerate}
\usepackage{multirow}
\usepackage{microtype}
\usepackage{mathptm}
\usepackage{todonotes}
\usepackage{listings}
\usepackage{xspace}
\usepackage{enumitem}

\IEEEoverridecommandlockouts

\begin{document}
\title{Efficient Verification of Multi-Property Designs (The Benefit of Wrong Assumptions)$^*$\thanks{\!\!\!\!$^*$
    This is an extended version of the paper published at DATE-2018.}\thanks{%
Supported by ERC project 280053 ``CPROVER'', the H2020 FET OPEN 712689 SC$^2$
and SRC contracts no.~2012-TJ-2269 and 2016-CT-2707.}}
\author{\IEEEauthorblockN{Eugene Goldberg, Matthias G{\"u}demann, Daniel Kroening}
\IEEEauthorblockA{Diffblue Ltd.\\
Oxford, UK}\and
\IEEEauthorblockN{Rajdeep Mukherjee}
\IEEEauthorblockA{University of Oxford\\
UK}}

\def \giveproofs {}

\maketitle

\begin{abstract}
We consider the problem of efficiently checking a set of safety
properties $P_1,\dots,P_k$ of one design.  We introduce a new approach
called \ja-verification, where \ja stands for ``Just-Assume'' (as
opposed to ``assume-guarantee'').  In this approach, when proving a
property $P_i$, one assumes that every property $P_j$ for $j \neq i$
holds.  The process of proving properties either results in showing
that $P_1,\dots,P_k$ hold without any assumptions or finding a
``debugging set'' of properties.  The latter identifies a subset of
failed properties that are the first to break.  The design behaviors
that cause the properties in the debugging set to fail must be fixed
first.  Importantly, in our approach, there is no need to prove the
assumptions used.  We describe the theory behind our approach and
report experimental results that demonstrate substantial gains in
performance, especially in the cases where a small debugging set
exists.

\end{abstract}


\section{Introduction}

The advent of powerful model checkers based on
\textsc{Sat}~\cite{bmc,ken03,ic3,pdr} has created a new wave of
research in property checking.  This research has been mostly focused
on algorithms that verify a single property for a given design.
However, in practice, engineers write many properties for one
design (sometimes hundreds and even thousands). This demands
efficient and scalable techniques for automatic verification of
multiple properties for one design.

More specifically, the problem we address is as follows.  We are given
a transition relation and a set of initial states, which specify the
design.  In addition, we are given a set of safety properties
$P_1,\ldots,P_k$ that are \textit{expected to hold}. (In
Section~\ref{sec:etf_props}, we consider the case where some
properties are expected to fail.) We want to check if every property
$P_i$ holds. If not, we want to have an efficient way to identify
failed properties that point to wrong design behaviors. (Thus,
identification of \ti{all} failed properties is not mandatory.)  One
way to solve this problem is to check whether the property $P\,:=\,P_1
\wedge \ldots \wedge P_k$ holds.  We will call $P$ the \ti{aggregate}
property.  If~$P$ holds, then all properties $P_i$ are proven.
Otherwise, the generated \textit{counterexample} (\cex{} for short)
identifies a subset of the failed properties, but no information is
gained about the remaining properties.  The latter can be verified by
removing the failed properties from the set, and re-iterating the
procedure with a new aggregate property.

In this paper, we study an alternative approach where the properties
$P_i$ are verified separately. We will refer to this approach as
\ti{separate verification} as opposed to \ti{joint verification} of a
set of properties. We will highlight three main reasons for our
interest in separate verification. First, we want to study
multi-property verification in the context of an \pdr-like model
checker~\cite{ic3}.  Such a model checker will benefit from separate
verification by generating proofs that take into account the specifics
of each property. Besides, a property $P_i$ is a weaker version of the
aggregate property $P$. Thus, proving $P_i$ should be easier than $P$.
Second, each property $P_i$ is an over-approximation of the same set
of reachable states. Therefore, inductive invariants of already proven
properties can be re-used.  The third reason is as follows. If all
$P_i$ are true, there is a common proof of this fact, namely a proof
that $P$ holds. However, if some $P_i$ properties fail, there may not
be one universal \cex that explains all failures if the latter are
property specific. Separate verification is more relevant in such a
context.

In this paper, we introduce a version of separate verification called
\ja-verification. Here, \ja{} stands for ``Just-Assume'', as opposed to
``assume-guarantee''.  In \ja-verification, one proves property $P_i$,
assuming that every other property $P_j$, for $j \neq i$, holds,
regardless whether it is true. We will call such a proof \ti{local} as
opposed to a \ti{global} proof that $P_i$ is true where no assumptions
are made. \ja-verification results in constructing the set of
properties $P_i$ that failed locally (if any). We show that if the
aggregate property $P$ fails, there is at least one property $P_i$
that fails \textit{both} globally and locally. Thus, if \emph{all}
properties $P_i$ hold locally it also means they hold globally.

If $P_i$ holds locally, it either holds globally as well or every \cex{}
that breaks $P_i$ fails some other property $P_j$ \ti{before} this
\cex{} fails $P_i$.  That is, a \cex{} for $P_i$ contains a shorter
\cex{} for $P_{j}$.  This suggests that if $P_i$ holds locally, its
failure (if any) is most likely \textit{caused by} failures of other
properties.  For this reason, we call the set of properties that
\ti{fail} locally a \ti{debugging set}.  This debugging set of
properties points to design behaviors that need to be fixed in the first
place.  The approach guarantees that the failure of a property from the
debugging set is not preceded by a failure of any other property.
\ja-verification constructs a debugging set as follows: when proving
$P_i$ it assumes that all $P_j$, for $j \neq i$, hold even when some of
them fail.  Thus, even the wrong assumption that all $P_j$, $j\neq i$
hold proves to be useful.  For that reason, we use the term ``Just
Assume'' to name our approach.

To improve the efficiency of \ja-verification, we exploit the fact
that \pdr proves a property by strengthening it to make this
property inductive.  Specifically, we show that the strengthening clauses
generated by \pdr can be re-used when making any other property
inductive if the same transition relation and initial states are
used. Thus, in \ja-verification, clauses generated to make $P_i$
inductive are re-used when proving $P_j$, $j \neq i$.

Our contribution is threefold. First, we describe a new method of
multi-property verification called \ja-verification (see
Section~\ref{sec:ja_verif}). It is based on the machinery of local
proofs (Sections~\ref{sec:lg_proofs} and~\ref{sec:debug}) and re-use
of strengthening clauses (Section~\ref{sec:reusing}).  Second, we show
that \ja-verification generates \cex{}s only for a special subset of
failed properties called a debugging set. This is very important since
computing a \cex{} can be quite expensive (e.g., if a counter is
involved). Third, we provide implementation details
(Section~\ref{sec:implem}) and experimental results showing the
viability of \ja-verification (Sections~\ref{sec:exper_jnt_sep},
\ref{sec:exper_with_sep},\ref{sec:parallel}). In particular, in
Section~\ref{sec:parallel}, we give evidence that \ja-verification
facilitates parallel computing.


\section{Local And Global Proofs}
\label{sec:lg_proofs}

Separate verification is based on the assumption that, in general,
proving $P_i$ is easier than $P_1 \wedge \ldots \wedge P_k$ because
$P_i$ is a weaker property. In this section, we discuss how one can
make a proof of a property simpler if this proof is only needed in the
context of proving a stronger property.

%
%
\subsection{Definitions}

We will denote the predicates of the \ti{transition relation} and the
\ti{initial states} as $T(S,S{\,'})$ and $I(S)$ respectively.  Here $S$
and $S'$ are sets of present and next state variables respectively.
An assignment \pnt{s} to variables $S$ is called a \ti{state}.  We
will refer to a state satisfying a predicate $Q(S)$ as a
$Q$-\ti{state}.  A~property is just a predicate $P(S)$.  We will say
that a $P$-state (respectively $\overline{P}$-state) is a \ti{good}
(respectively \ti{bad}) state.  A~property $P$ is called
\ti{inductive} with respect to $T$ if $P \wedge T \rightarrow P{\,'}$
holds where a primed predicate symbol means that the predicate in
question depends on next state variables $S{\,'}$.

We will call a sequence of states \tr{s}{j}{k} a \ti{trace} if
$T(\pnt{s_i},\pnt{s_{i+1}})$ is true for $i=j,\ldots,k-1$.  We will
call the trace above \ti{initialized} if \pnt{s_j} is an $I$-state.
Given a property $P(S)$ where $I \rightarrow P$, a \cex{} is an
\emph{initialized trace} \tr{s}{0}{k} where \pnt{s_i}, $i=0,\ldots,k-1$
are $P$-states and \pnt{s_k} is a $\overline{P}$-state. We will refer
to a state transition system with initial states $I$ and transition
relation $T$ as an $(I,T)$-\ti{system}.  Given an $(I,T)$-system,
checking a property $P$ is to find a \cex{} for $P$ or to show that
none exists.

%
%
\subsection{Hardness of proving strong and weak properties}

Let $P$ and $Q$ be two properties where $Q$ is weaker than $P$, i.e.,
$P \rightarrow Q$. On the one hand, verification of $Q$ should be easier
because one needs to prove unreachability of a smaller set of bad
states. On the other hand, $P$ can be inductive even if $Q$ is not.  In
fact, the essence of \pdr{} is to turn a non-inductive property into an
inductive one by adding strengthening clauses. This makes the modified
property easier to prove despite the fact that it is stronger from a
logical point of view.

The reason for the paradox above is as follows. The set of traces one
needs to consider to prove $Q$ is not a subset of those one considers
when proving $P$. To prove $P$, one needs to show that there is no
initialized trace of $P$-states leading to a
$\overline{P}$-state. Thus, one does not consider traces where two
$\overline{P}$-states occur. Proving $Q$ is reduced to showing that
there is no initialized trace of $Q$-states leading to a
$\overline{Q}$-state. Since $P \rightarrow Q$, a $\overline{Q}$-state
is a $\overline{P}$-state as well. On the other hand, a $Q$-state can
also be a $\overline{P}$-state. Thus, in contrast to the case when
we prove $P$, to prove $Q$ one has to consider traces that may include
two or more different $\overline{P}$-states.

We will refer to a regular proof of $Q$ (where one shows that no
initialized trace of $Q$-states leads to a $\overline{Q}$-state) as a
\ti{global} one. In the next subsection, we discuss reducing the
complexity of proving $Q$ using the machinery of local proofs.

%
%
\subsection{Local proofs}
\label{ssec:lproofs}

The intuition behind local proofs is as follows. Suppose that one
needs to prove a property $Q$ as a step in proving a stronger property
$P$. Then it is reasonable to ignore traces that do not make sense
from the viewpoint of proving $P$. Proving $Q$ \ti{locally in the
  context of} $P$, or just \ti{locally} for short, is to show that
there does not exist an initialized trace of $P$-states (rather than
$Q$-states) leading to a $\overline{Q}$-state.

The importance of local proofs is twofold. First, to prove $Q$
locally, one needs to consider only a subset of the traces to prove
$P$ (because the set of $\overline{Q}$-states is a subset of that of
$\overline{P}$-states).  Thus, in terms of the set of traces to
consider, a weaker property becomes also ``easier''. Second, as we
show in Section~\ref{sec:ja_verif}, to prove the aggregate property $P
:= P_1 \wedge \ldots \wedge P_k$, it suffices to prove all properties
$P_i$ \ti{locally} with respect to $P$.

It is convenient to formulate the notion of a local proof in terms of
a modified transition relation $T$. We will call this modification
\ti{the projection of} $T$ \ti{onto property} $P$ and denote it as
$T^P$. It is defined as follows.
\begin{itemize}
\item  $T^P(\pnt{s},\pnt{s'}) = T(\pnt{s},\pnt{s'})$, if \pnt{s} is a $P$-state.
\item $T^P(\pnt{s},\pnt{s'}) = 0$ if \pnt{s} is a $\overline{P}$-state and $\pnt{s} \neq \pnt{s'}$.
 \item $T^P(\pnt{s},\pnt{s'}) = 1$ if \pnt{s} is a $\overline{P}$-state and $\pnt{s} = \pnt{s'}$.
\end{itemize}

Informally, $T^P$ is obtained from $T$ by excluding any transitions
from a $\overline{P}$-state other than a transition to itself.  Hence,
a trace in $(I,T^P)$-system cannot have two different
$\overline{P}$-states. Thus, a local proof of $Q$ with respect to
property $P$, as we introduced above, is just a regular proof with
respect to $T^P$. (In turn, proving $Q$ globally is done with respect
to $T$.)

\begin{proposition}
\label{prop:reuse_inv}
Let $P$ be inductive with respect to transition relation $T$.  Then
any property $Q$ weaker than $P$ (i.e., $P \rightarrow Q$) is
inductive with respect to $T^P$.
\end{proposition}
\ifdefined \giveproofs 
\begin{mproof}
  Assume the contrary. That is $Q$ is not inductive with respect to
  $T^P$ and hence $Q \wedge T^P \rightarrow Q'$ does not hold. Then
  there is a transition (\pnt{s},\,\pnt{s'}) such that
  \begin{itemize}
  \item $T^P(\pnt{s},\pnt{s'}) = 1$ and
  \item $Q(\pnt{s}) = 1$ and $Q(\pnt{s'}) = 0$ (and hence $\pnt{s} \neq \pnt{s'}$)
  \end{itemize}
  Since $P \rightarrow Q$, then $P(\pnt{s'}) = 0$ as well. Since
  $\pnt{s} \neq \pnt{s'}$ and $T^P(\pnt{s},\pnt{s'})~=~1$, from
  definition of $T^P$ it follows that $P(\pnt{s}) = 1$ and
  $T(\pnt{s},\pnt{s'})=1$.  So (\pnt{s},\,\pnt{s'}) is a transition
  from a $P$-state to a $\overline{P}$-state allowed by $T$.  Then $P$
  is not inductive with respect to $T$. We have a contradiction.
\end{mproof}
\fi

Proposition~\ref{prop:reuse_inv} states that in terms of proofs by
induction, proving $Q$ locally with respect to a stronger property $P$
is \ti{at most} as hard as proving $P$ itself.


\section{Local Proofs And Debugging}
\label{sec:debug}

In this section, we explain how the machinery of local proofs can be
used to address the following problem.  Given a property~$P$ that
failed, find a weaker property $Q$ that is false as well and can be
viewed as an \ti{explanation} for failure of $P$.  The subtlety here
is that not every failed property $Q$ where $P \rightarrow Q$ can be
viewed as a reason for why $P$ fails.  We will refer to this problem
as the \ti{debugging} problem.

To address the debugging problem one first needs to clarify the
relation between local and global proofs.
\begin{proposition}
  \label{prop:lg_proofs}
  Let $P \rightarrow Q$.
  \begin{enumerate}[label=\Alph*)]
\item If property $Q$ holds with respect to transition relation $T$
  (i.e., globally), it also holds with respect to $T^P$ (i.e.,
  locally).
  \end{enumerate}
  The opposite is not true.
  \begin{enumerate}[label=\Alph*),resume] 
\item   If $Q$ holds with respect
to $T^P$, it either holds with respect to $T$ or it fails with respect
to $T$ and every \cex{} contains at least two $\overline{P}$-states
\pnt{s_i} and \pnt{s_j} where $\pnt{s_i} \neq \pnt{s_j}$.
\end{enumerate}
\end{proposition}
\ifdefined \giveproofs 
\begin{mproof}
  A) Assume that $Q$ does not hold with respect to $T^P$. Then there
  is an initialized trace of $P$-states leading to a
  $\overline{Q}$-state. Since $T^P$ inherits all transitions of $T$
  from $P$-states, this trace is valid with respect to $T$. Since $P
  \rightarrow Q$, every $P$-state is also a $Q$-state. So there is an
  initialized trace of $Q$-states leading to a $\overline{Q}$-state
  that is valid with respect to $T$. Hence $Q$ does not hold with
  respect to $T$ and we have a contradiction.

  B) Assume the contrary, i.e. $Q$ holds with respect to $T^P$ and
  there is a \cex with respect to $T$ containing only one
  $\overline{P}$-state. (So neither $Q$ holds globally nor every \cex
  contains at least two $\overline{P}$-states). The \cex above is also
  a \cex with respect to $T^P$ and hence $Q$ fails locally. So we have
  a contradiction.
\end{mproof}
\fi

Informally, Proposition~\ref{prop:lg_proofs} means that proving $Q$
locally is ``as good as'' proving globally modulo \cex{}s that do not
make sense from the viewpoint of proving $P$.  These \cex{}s have at
least two $\overline{P}$-states.

One can use Proposition~\ref{prop:lg_proofs} for solving the debugging
problem as follows. Suppose that $Q$ does not hold locally. This means
that there is a \cex{} of $P$-states leading to a
$\overline{Q}$-state. Since $P \rightarrow Q$, a $\overline{Q}$-state
is a $\overline{P}$-state as well. So this \cex{} is also a regular \cex{}
for $P$. In other words, the fact that $Q$ fails locally means that
$Q$ can be viewed as a reason for failure of $P$.

Suppose that $Q$ holds locally. Assume that $Q$ fails globally and
\tr{s}{0}{m} is a \cex{} where \pnt{s_m} is a $\overline{Q}$-state. From
Proposition~\ref{prop:lg_proofs}, it follows that this \cex{} has at
least two $\overline{P}$-states. One of these states is \pnt{s_m}
(because $P \rightarrow Q$). Another $\overline{P}$-state is one of
$Q$-states \pnt{s_i}, $i=1,\ldots,m-1$. This means that failure of $Q$
is not a reason for failure of $P$. Indeed, in every \cex{} for $Q$,
property $P$ fails \ti{before} $Q$ does.

Summarizing, if property $Q$ fails (respectively holds) locally with
respect to $P$, failure of $Q$ is a reason (respectively cannot be a
reason) for failure of $P$.


\section{\ja-verification}
\label{sec:ja_verif}

In this section, we present a version of separate verification called
``Just-Assume'' or \ja-verification. As before, $P$ denotes the
aggregate property $P_1 \wedge \ldots \wedge P_k$ and $T^P$ denotes the
projection of $T$ onto $P$ (see Subsection~\ref{ssec:lproofs}).  Since
every property $P_i$ is a weaker version of $P$, one can use the results
of Sections~\ref{sec:lg_proofs} and~\ref{sec:debug} based on the
machinery of local proofs.

We now provide a justification of proving weaker properties locally in
the context of multi-property verification. By using the transition
relation $T^P$ to prove $P_i$, one essentially assumes that every
property $P_j$, $j \neq i$ holds. While this may not be the case,
nevertheless it works for two reasons.  The first reason is that if
the aggregate property $P$ fails, there is a time frame where $P$ (and
hence some property $P_i$) fails for the first time. Let this be time
frame number $m$. For every time frame number $p$ where $p < m$, the
assumption that every property $P_j$, $j \neq i$ holds \ti{is true}.
Thus, if $P$ fails, there is at least one property (in our case $P_i$)
that fails even with respect to $T^P$.

Here is the second reason why assuming $P_j$, $j \neq i$ works.
To get some debugging information when proving property
$P_i$, one is interested in traces where $P_i$ fails \ti{before} any
other property does. By assuming $P_j$, $j \neq i$ is true, one drops
the traces where $P_i$ fails \ti{after} some $P_j$, $j \neq i$ has
failed.

The propositions below formalize the relation between local proofs and
multi-property verification.

%
%
\begin{proposition}
\label{prop:joint_sep}
  Property $P$ holds with respect to $T$ iff
  every $P_j$, $j=1,\ldots,k$  holds with respect to $T$.
\end{proposition}
\ifdefined \giveproofs 
\begin{mmproof}
  \ti{If part.} Let $P_j$, $j=1,\ldots,k$ hold with respect to $T$.
  Assume the contrary i.e. $P$ does not hold. Then there is a \cex{}
  \tr{s}{0}{m} where $\pnt{s_i},i=0,\ldots,m-1$ are $P$-states and
  \pnt{s_m} is a $\overline{P}$-state. This means that \pnt{s_m}
  falsifies some property $P_j$. Since a $P$-state is also a
  $P_j$-state, the \cex{} above is an initialized trace of
  $P_j$-states leading to a $\overline{P_j}$ state.  Hence $P_j$ does
  not hold and we have contradiction.

  \ti{Only if part.} Let $P$ hold with respect to $T$. Assume the
  contrary i.e. a property $P_j$ does not hold. Then there is a \cex{}
  \tr{s}{0}{m} where $\pnt{s_i}$, $i=0,\ldots,m-1$ are $P_j$-states
  and \pnt{s_m} is a $\overline{P_j}$-state. Since $P \rightarrow P_j$
  holds, \pnt{s_m} is a $\overline{P}$-state as well.  As far as
  states $\pnt{s_i}$, $i=0,\ldots,m-1$ are concerned one can have the
  following two situations.
  \begin{itemize}
  \item All these states are $P$-states as well. Then the \cex{} above breaks
    $P$ and we have a contradiction.
  \item At least one state \pnt{s_i}, $0 \leq i \leq m-1$ is
    a $\overline{P}$-state. In this case, trace \tr{s}{0}{i}
    is a \cex{} breaking $P$ and we have a contradiction.~~~QED
  \end{itemize}
\end{mmproof}
\fi
\begin{proposition}
\label{prop:glob_loc}
  Property $P$ holds with respect to $T$ iff
  $P$ holds with respect to $T^P$.
\end{proposition}
\ifdefined \giveproofs 
\begin{mproof}
  Any trace of $P$-states leading to a $\overline{P}$-state is valid
  both with respect to $T$ and $T^P$. So if $P$ fails with respect to $T$
  it also does with respect to $T^P$ and vice versa.
\end{mproof}
\fi

%
%

\begin{proposition}
\label{prop:glob_to_loc}
  Property $P$ holds with respect to $T$ iff
  every $P_i$, $i=1,\ldots,k$ holds with respect to $T^P$.
\end{proposition}
\ifdefined \giveproofs 
\begin{mproof}
  From Proposition~\ref{prop:glob_loc} it follows that $P$ holds with
  respect to $T$ iff it holds with respect to $T^P$. After replacing
  $T$ with $T^P$ in Proposition~\ref{prop:joint_sep}, one concludes
  that $P$ holds with respect to $T^P$ iff every $P_i$, $i=1,\ldots,k$
  holds with respect to $T^P$.
\end{mproof}
\fi

%
%
We will refer to the subset of \s{P_1,\ldots,P_k} that consists of
properties that fail with respect to $T^P$, i.e., locally as a
\ti{debugging set}. The following proposition justifies this name.
\begin{proposition}
  Let the aggregate property $P$ fail.  Let $D$ denote the debugging set
  of properties. Then the failure of properties of $D$ is the reason for
  the failure of $P$ in the following sense. For each \cex{}
  \tr{s}{0}{m} for property $P$, the state \pnt{s_m} that falsifies $P$
  also falsifies at least one property $P_i \in D$.
\end{proposition}
\ifdefined \giveproofs 
\begin{mproof}
  Assume the contrary, i.e., there is a \cex{} for $P$ such that
  \pnt{s_m} falsify only properties of \s{P_1,\ldots,P_k} that are not
  in $D$.  Let $P_i$ be a property falsified by \pnt{s_m}. Since this
  \cex{} consists of $P$-states leading to a $\overline{P}_i$-state,
  then $P_i$ fails locally and so it is in $D$. Thus, we have a
  contradiction.
\end{mproof}
\fi

\lstdefinestyle{base} {
  basicstyle = \ttfamily\footnotesize,
  moredelim=**[is][\color{blue}]{~}{~},
}
\begin{example}
\label{example}
  The Verilog code below gives an example of an 8-bit counter. This
  counter increments its value every time the \ti{enable} signal is
  true. Once the counter reaches the value of \ti{rval} it resets its
  value to 0. We also want to reset the counter when signal \ti{req}
  is true (\ti{regardless} of the current value of the counter).
  However, the code contains a buggy line (marked in blue), which
  prohibits a reset \ti{only} unless \ti{req} is true.
\begin{lstlisting}[language=Verilog,style=base] 
module counter(enable,clk,req);
   parameter  rval = 1 << 7;
   input      enable, clk, req;
   reg [7:0]  val;
   wire       reset;

   initial val = 0;

   ~assign reset = ((val == rval) && req); ~

   always @(posedge clk) begin
      if (enable) begin           
         if (reset) val = 0;
         else val = val+1;
      end           
   end

   P0: assert property (req == 1);
   P1: assert property (val <= rval);
   
endmodule
\end{lstlisting}
\end{example}

Let us consider verification of properties $P_0$ and $P_1$ specified by the
last two lines of the module {\tt counter}.  Property $P_0$ fails globally
in every time frame because \ti{req} is an input variable taking values 0
and 1.  Property $P_1$ fails globally due to the bug above.  Note however,
that only property $P_0$ fails \ti{locally}.  Indeed, $P_0$ fails even under
assumption $P_1 \equiv 1$.  However, $P_1$ becomes true if one assumes $P_0
\equiv 1$.  The latter means that $\mi{req} \equiv 1$ and so the counter
always resets on reaching value \ti{rval}.  So the debugging set consists
only of $P_0$.  The fact that $P_1$ holds locally means that either $P_1$ is
true globally or that any \cex failing $P_1$ first fails $P_0$.  The latter
implies that the failure of $P_1$ is caused by incorrect handling of
variable \ti{req}.

%
%
\begin{table}
\small
\caption{\ti{Example with a counter. Time limit is 1 hour}}
\vspace{-5pt}
\scriptsize
\begin{center}
\begin{tabular}{|r|r|r|r|r|c|} \hline
\#bits      & \multicolumn{4}{c|}{solving globally} &solving     \\ \cline{2-5}
            & \multicolumn{2}{c|}{\Abc (bmc)}&\multicolumn{2}{c|}{\Abc (pdr)} & \mbox{locally}\\ \cline{2-5}
            & \#time frames& time &\#time frames & time   &                       \\ \hline
  8         &  128         & 0.3\,s & 10       & 0.1\,s   &   0.01\,s       \\ \hline
  12        &  2,048       & 723\,s & 51       & 1.7\,s   &   0.02\,s       \\ \hline
  14        &   $*$        &  $*$   & 118      & 9.9\,s   &   0.02\,s       \\ \hline
  16        &   $*$        &  $*$   & 269      & 113\,s   &   0.02\,s       \\ \hline
  18        &   $*$        &  $*$   & 315      & 1,278\,s &   0.02\,s       \\ \hline
  20        &   $*$        &  $*$   & $*$      & $*$      &   0.02\,s       \\ \hline

\end{tabular}                
\end{center}
\vspace{-15pt}
\label{tbl:counter}
\end{table}

Note that proving $P_1$ false globally is hard for a large counter because a
\cex consists of all states of the counter from 0 to \ti{rval}.  On the
contrary, proving $P_1$ true under assumption $P_0 \equiv 1$ is trivial
because $P_1$ is inductive under this assumption.  In
Table~\ref{tbl:counter}, we compare proving properties $P_0$ and $P_1$ above
globally and locally.  The first column gives the size of the counter.  The
next four columns give the results of solving $P_0$ and $P_1$ globally by
\Abc, a mature tool developed at UC~Berkeley~\cite{abc}.  The first pair of
columns gives the results of Bounded Model Checking~\cite{bmc} (the largest
number of used time frames and run time).  The next pair of columns provides
results of \textsc{Pdr} (i.e., \pdr).  Finally, we give the results of
solving $P_0$ and $P_1$ locally by our tool (see Section~\ref{sec:implem}).

The results show that bounded model checking soon becomes impractical, as the
number of time frames increases exponentially.  \textsc{Abc}'s \textsc{Pdr}
solves more cases, but to generate a \cex, it has to consider a quickly
increasing number of time frames as well.  For \ja-verification, the size of
the counter has no influence on the run time.  While the counter is a purely
synthetic example, in practice, one often has to find \mbox{so-called}
\emph{deep} counterexamples.  A~system with complex inner state might
require a long sequence of steps to reach a buggy state.




\section{Handling Properties Expected To Fail}
\label{sec:etf_props}

When proving properties $P_i$, $i=1,\dots,k$ in \ja-verification,
as introduced in Section~\ref{sec:ja_verif}, one excludes the traces
where a property $P_j$, $j \neq i$ fails before $P_i$ does.  This is
based on the assumption that the properties that are the first to fail
indicate design behaviors that need to be fixed first.
However, this assumption is unreasonable when a property
$P_j$ that fails before $P_i$ is \ti{Expected To Fail}
(\textsc{Etf}). For instance, to ensure that a state \pnt{s} is
reachable, one may formulate an \textsc{Etf} property $P_j$ where
\pnt{s} is a $\overline{P}_j$-state. In this case, excluding the
traces where $P_j$ fails before $P_i$ is a mistake.

One can easily extend \ja-verification to handle \textsc{Etf} properties as
follows.  Suppose that our objective is to prove every property $P_i$ that
is \emph{Expected To Hold} (\textsc{Eth}).  In addition, for every
\textsc{Etf} property we want to find a \cex{} that does not break any
\textsc{Eth} property.  Then, to solve $P_i$, $i=1,\dots,k$ \ti{locally} one
assumes that every \textsc{Eth} $P_j$, $j\neq i$ is true.  Thus, we exclude
the traces where \textsc{Eth} properties fail before $P_i$, even if the
latter is an \textsc{Etf} property.

\section{\pdr{} And Clause Re-using}
\label{sec:reusing}
So far, we discussed the machinery of local proofs without specifying
the algorithm used to prove a property. In this section, we describe
an optimization technique applicable if property checking is performed
by \pdr{}~\cite{ic3}. The essence of this technique is to re-use
strengthening clauses generated by \pdr{} for property $P_i$ to
strengthen another property $P_j$, $j \neq i$. Before describing
clause re-using, we give a high-level view of \pdr.

%
%
\subsection{Brief description of \pdr{}}
Let $Q$ be a property of an $(I,T)$-system where $I \rightarrow Q$.
If $Q$ holds, there always exists a predicate $G(S)$ such that
$Q\wedge G$ is inductive with respect to $T$.  Then $(Q \wedge G
\wedge T) \rightarrow (Q' \wedge G')$. (Recall that a primed predicate
symbol means that the predicate in question depends on next state
variables $S'$.) The fact that $Q \wedge G$ is inductive implies that
$Q \wedge G$ is an over-approximation of the set of states reachable
in the $(I,T)$-system in question. Therefore, for every state \pnt{s}
reachable in $(I,T)$-system, $Q(\pnt{s}) \wedge G(\pnt{s}) = 1$.

Let $F$ denote $Q \wedge G$ i.e property $Q$ strengthened by $G$.  In
\pdr{}, formulas are represented in conjunctive normal form and the
predicate $F$ is constructed as a set of \ti{clauses} (disjunctions of
literals). Let \rch{I}{T}{j} denote the set of states reachable from
$I$-states in at most $j$ transitions. To construct the formula $F$,
\pdr{} builds a sequence of formulas $F_0,\dots,F_m$ where $F_0 = I$
and $F_j, j=1,\dots,m$ specifies an over-approximation of
\rch{I}{T}{j}. That is, if a state \pnt{s} is in \rch{I}{T}{j}, then
\pnt{s} satisfies $F_j$, i.e., $F_j(\pnt{s}) = 1$. A formula $F_j$,
$j> 0$ is \ti{initialized} with $Q$. Then $F_j$ is strengthened by
adding so called \emph{inductive clauses}. The objective of this
strengthening is to exclude the $F_j$-states from which a bad state is
reachable in one transition.  The exclusion of an $F_j$-state may
require excluding $F_i$-states, $i < j$ (by adding inductive clauses
to $F_i$) from which a bad state can be reached in $j-i+1$
transitions. If $Q \wedge F_i$ becomes inductive for some value of $i,
i \leq j$, property $Q$ holds. Clause $C$ is called \ti{inductive
  relative to $F_j$} if $I \rightarrow C$ and $C \wedge F_j \wedge T
\rightarrow C'$ hold.  In this case, every $\pnt{s} \in \rch{I}{T}{j}$
satisfies $C$ and so $F_j \wedge C$ is still an over-approximation of
\rch{I}{T}{j}.

%
%
%
\subsection{Re-using strengthening clauses}
\label{ssec:reusing}
The idea of re-using strengthening clauses is based on the following
observation. Suppose that $G_Q$ is a set of clauses which makes $Q$
inductive in the $(I,T)$-system. This means that $Q \wedge G_Q$ is an
over-approximation of the set of all states reachable in the
$(I,T)$-system. Hence a state $\pnt{s} \in \rch{I}{T}{j}$ satisfies
$G_Q$, for any value $j \ge 0$.  Suppose one needs to prove some other
property $R$. Then, when constructing a formula over-approximating
\rch{I}{T}{j}, one can initialize this formula with $R \wedge G_Q$,
rather than with $R$.

%
%
\subsection{State lifting in \pdr{}}
\label{ssec:lifting}
In this subsection, we briefly describe an important technique of \pdr
called \emph{state lifting}~\cite{pdr,lifting}. This description is
used in Section~\ref{sec:implem} when explaining how local proofs are
implemented.  Let \pnt{s} be a state from which a bad state is
reachable. \pdr{} tries to exclude \pnt{s} by generating an inductive
clause falsified by \pnt{s}.  If such a clause cannot be built
immediately, \pdr{} tries to exclude every state \pnt{q} of the
previous time frame from which there is a transition to state
\pnt{s}. The number of those states \pnt{q} can be very large. So,
\pdr tries to ``lift'' \pnt{q} to a cube \cof{C}{q} that not only
contains the state \pnt{q} itself, but many other states that are one
transition away from state \pnt{s}. Subsequently, \pdr tries to
exclude all states of \cof{C}{q} in one shot. Informally, the larger
\cof{C}{q}, the greater the performance boost by lifting.


\section{Implementation}
\label{sec:implem}
We use a version of \pdr developed in our research group.  We will refer
to it as \db where ``db'' stands for Diffblue. \db uses the front-end of
\textsc{ebmc}~\cite{mkm2015}. We will refer to our
implementation of \ja-verification based on \db as \Sep.  The latter is a
Perl script that calls \db in a loop for proving individual properties.

\subsection{Using properties as constraints}
\label{ssec:prop_constr}
Let $P_1,\ldots,P_k$ be the set of properties to be proved.  Proving
$P_i$ locally means showing that there is no initialized trace of
$P$-states that leads to a $\overline{P_i}$-state, where $P$ is the
aggregate property $P_1 \wedge \ldots \wedge P_k$.  To guarantee that
all present states satisfy~$P$, \db adds constraints to the transition
relation $T$ that force $P_j$, $j \neq i$ to be equal to~1.  Adding
constraints to $T$ affects the lifting procedure of \pdr (see
Subsection~\ref{ssec:lifting}).  The reason is that one needs to
guarantee that all states of the cube \cof{C}{q} obtained by lifting a
state \pnt{q} satisfy the constraints in question.  In our case, all
states of \cof{C}{q} must be $P$-states.  Unfortunately, this can
drastically decrease the size of \cof{C}{q} and therefore reduce the
effectiveness of lifting. For that reason, \db has an option to make
the lifting procedure ignore the constraints forcing $P_j, j \neq i$
to be equal 1.

The relaxation of lifting above can lead to appearance of ``spurious''
{\cex}s that contain transitions from $\overline{P}$-states to other
states. (This can occur only if $P$ does not hold).  If a spurious
\cex is generated when proving property $P_i$, then \db is invoked again.
This time the lifting procedure is forced to respect the constraints
specified by $P_j, j \neq i$.

\subsection{Implementation of clause re-using}
\label{ssec:impl_reuse}
The correctness of re-using strengthening clauses is discussed in
Subsection~\ref{ssec:reusing}. Assume that properties $P_1,\ldots,P_k$
are processed in the order they are numbered.  Let \Sub{G}{P_1} denote
the strengthening of property $P_1$, i.e., $P_1 \wedge \Sub{G}{P_1}$ is
inductive with respect to $T$. \Sep maintains an external file
\clauseDB that collects strengthening clauses. Therefore, after making
$P_1$ inductive, the clauses of \Sub{G}{P_1} are written to \clauseDB{}.

When \db is invoked to prove $P_2$, the clauses of \Sub{G}{P_1} are
extracted from {\clauseDB}. When proving $P_2$, the formula
over-approximating \rch{I}{T}{j} is initialized with $P_2 \wedge
\Sub{G}{P_1}$ (rather than with $P_2$). Let \Sub{G}{P_2} denote the
strengthening clauses added to $P_2 \wedge \Sub{G}{P_1}$ to make the
latter inductive with respect to $T$. These clauses are written to
\clauseDB{} (that already contains \Sub{G}{P_1}). In general, when \db
is invoked to prove $P_j$, all clauses $\bigcup_{i<j}\Sub{G}{P_i}$ are
extracted from \clauseDB to be used in the proof.


%
%
\section{Comparing  local and global proofs}
\label{sec:comp_lg}
In Sections~\ref{sec:exper_jnt_sep},~\ref{sec:exper_with_sep}
and~\ref{sec:parallel}, we experimentally compare methods based on
local and global proofs.  In this section, we summarize the
information we provided earlier to help better understand these
experimental results. We will refer to methods proving properties
locally (respectively globally) as a local (respectively global)
approach.

Let \mb{P} denote the set \s{P_1,\dots,P_k} of properties to
verify. There are two cases where local and global approaches provide
the same information. Assume one managed to prove every property of
\mb{P} locally.  In this case, every property of \mb{P} holds globally
as well. Thus global and local approaches provide the same information
here. Now assume that one proved $P_i \in \mb{P}$ false locally (and
hence globally).  Assume also that in a global approach one proved
$P_i$ false and in the generated \cex property $P_i$ is \ti{the first}
to fail. Then, $P_i$ is ``inadvertently'' proved false not only
globally but locally too. In this case, both local and global
approaches provide the same information for $P_i$.

Now consider the cases where local and global approaches provide
different information for a failed property. Assume that one proved
$P_i$ false globally and the generated \cex falsifies at least one
property $P_j \in \mb{P}$, $i \neq j$ before $P_i$. Consider the
following two cases. The first case is that one proves $P_i$ false
locally. Then the local approach provides more information than its
global counterpart because in addition to proving $P_i$ false is shows
that $P_i$ is in the debugging set. The second case is that one proves
$P_i$ \ti{true} locally. In this situation, local and global
approaches do not have a clear winner. On the one hand, the global
approach gives more information by finding a \cex for $P_i$. On the
other hand, the local approach provides more information by showing
that \ti{every} \cex falsifying $P_i$ (if any) first falsifies some
other property of \mb{P}. Note that from the viewpoint of debugging,
the information provided by the local approach in the second case is
more useful since $P_i$ is shown not to be in the debugging set.

Finally, lets us consider the following situation. Let $\mb{P\,'}
\subset \mb{P}$ be the subset of properties proved true locally and
$\mb{P\,''}=\mb{P} \setminus \mb{P\,'}$. A property is in \mb{P\,''}
if it is proved false locally or it is too hard to solve or it has not
been tried yet. Since $\mb{P\,''} \neq \emptyset$, the fact that $P_i
\in \mb{P\,'}$ holds locally does not mean that it holds globally as
well. Thus, proving $P_i$ true globally provides more information than
in the local approach. However, from the debugging point of view,
proving $P_i$ true locally is almost as good as globally. This proof
means that $P_i$ cannot fail ``on its own''.  So one first needs to
focus on properties of \mb{P\,''}, in particular, to fix the design
behaviors causing property failures.


\section{\ja-verification Versus Joint Verification}
\label{sec:exper_jnt_sep}
In this section, we experimentally compare \ja-verification and joint
verification to show the viability of our approach to multi-property
verification. We use benchmarks from the multi-property track of the
\textsc{Hwmcc}-12 and 13 competitions. As we mentioned in
Section~\ref{sec:implem}, \ja-verification is implemented as a Perl
script \Sep that calls \db to process individual properties
sequentially.  In this paper, we do not exploit the possibility to
improve \ja-verification by processing properties in a particular
order.\footnote{A rule of thumb here is to verify easier properties
  first to accumulate strengthening clauses and use them later for
  harder properties.} Properties are verified in the order they are
given in the design description.  Joint verification is also
implemented as a Perl script called \Jnt, where \db is called to
verify the aggregate property $P := P_1 \wedge \ldots \wedge P_k$.
If~$P$ fails, the individual properties refuted by the generated \cex
are reported false.  \Jnt forms a new aggregate property by conjoining
the properties $P_i$ that are unsolved yet and calls \db again. This
continues until every property is solved.

Ideally, we would like to use designs with as many properties as
possible. However, for a design with a very large number of
properties, \ja-verification usually outperforms joint verification
(see Subsection~\ref{ssec:large}).  This problem can be addressed by
partitioning $P_1,\dots,P_k$ into smaller clusters of
properties~\cite{mprop_date11}, which is beyond the scope of this
paper. To make joint verification more competitive, in
Subsections~\ref{ssec:fail} and~\ref{ssec:hold}, we picked sixteen
designs (eight designs per subsection) that have less than a thousand
properties. For these designs, we cross-checked the results of \db in
joint verification with those reported by the latest version of
\Abc~\cite{abc}.\footnote{Joint verification is the natural mode of
  operation for \Abc. However, in contrast to \Jnt, \Abc does not
  re-start when a property is proved false and goes on with solving
  the remaining properties.}

The time limit for joint verification was set to 10 hours.  The time
limit used by \db in \ja-verification to prove one property is
indicated in the tables of results.  If~a property of a benchmark was
not solved by \db, the time limit was added to the total time of
solving this benchmark.  Unfortunately, the \textsc{Hwmcc}
competitions do not identify properties of multi-property benchmarks
that are expected to fail, if any (see
Section~\ref{sec:etf_props}). So, in the experiments we just assumed
that every property was expected to hold.

\subsection{A few designs with a large number of properties}
\label{ssec:large}

In this subsection, we compare \ja-verification with joint
verification on a few benchmarks that have a very large number of
properties. The point we are making here is that for such benchmarks,
\ja-verification is typically more robust than joint verification.
One of the reasons for that is as follows. When one conjoins a set of
very different properties joint verification may fail to prove the
aggregate property even if all properties are simple individually. For
instance, it may be the case that each property of this set depends on
a small subset of state variables and hence can be easily proved
separately. However, if different properties depend on different
subsets of variables, the aggregate property depends on a large subset
of state variables and becomes very hard to prove. Another reason for
the poor performance of joint verification is that the presence of a
few too-hard-to-solve properties $P_i$ can blow up the complexity of
the aggregate property $P$.

%
%
%
\begin{table}[h]
\small
\caption[caption]{\ti{A few designs with a large number of properties}}
\vspace{-10pt}
\scriptsize
\begin{center}
  \begin{tabular} {|p{15pt}|p{17pt}|p{14pt}|p{14pt}|p{15pt}|p{14pt}|p{14pt}|p{14pt}|p{22pt}|} \hline
            name        &\!\!\#all\!\! & \!\!\#props\!\!& \multicolumn{2}{c|}{Joint verification} & \multicolumn{3}{c|}{~~~\ja-verification~~~~~~~~~~}\\ \cline{4-8}
            &  props & tried &\mbox{\#un-}&     time     & time & \mbox{\#un-}      & time    \\ 
            &        &       & solved         &              & limit &  solved     &         \\ \hline
 6s400     & 13,784 & 100   & 100    &   10\,h         & 0.3\,h    &\tb{3}  & \tb{3,167\,s}   \\ \hline     
 6s355     & 13,356 & 100   & 100    &   10\,h         & 0.3\,h    &\tb{2}  & \tb{2,175\,s}    \\ \hline     
           &        & 100   & 0      &   2,817\,s      & 0.3\,h    & 0      & \tb{1,974\,s}    \\ \cline{3-8}
 6s289     & 10,789 & 200   & \tb{0} &   \tb{1,095\,s} & 0.3\,h    & 2      &  1.1\,h          \\ \cline{3-8}
           &        & 500   & 100    &   10\,h         & 0.3\,h    & \tb{3}   & \tb{2.1\,h}      \\ \hline
           &        & 100   & 0      &   \tb{919\,s}   & 0.3\,h    & 0      & 1,126\,s        \\ \cline{3-8}
 6s403     &  2,382 & 200   & \tb{0} &   \tb{828\,s}   & 0.3\,h    & 1      & 2,329\,s         \\ \cline{3-8}
           &        & 500   & \tb{0} &   \tb{717\,s}   & 0.3\,h    & 1      & 3,265\,s         \\ \hline

\end{tabular}
\end{center}
\label{tbl:many_props}
\end{table}

%

In this experiment, we used \ja-verification and joint verification to
verify the first $k$ properties of a benchmark. (Joint verification was
performed by \db). The results of the experiment are given in
Table~\ref{tbl:many_props}.  The value of $k$ is given in the third
column. The first and second columns of this table provide the name of
a benchmark and the total number of properties.  For benchmarks 6s400 and
6s355, \ja-verification clearly outperformed joint verification. For
benchmark 6s289 both \ja-verification and joint verification performed
well for $k=100$ and $k=200$. However, for $k=500$ \ja-verification
outperformed joint verification. Benchmark 6s403 was the only one out
of  four where joint verification outperformed \ja-verification.

%
%
\begin{table}
\small
\caption{\ti{Designs with failed properties. Many properties of 6s258,
    6s207, 6s254, 6s335, and 6s380 are false globally in joint
    verification but true locally in \ja-verification. `mem' means
    running out of memory}}
\vspace{-15pt}
\scriptsize
\begin{center}
\begin{tabular}{|p{13pt}|@{\quad}r@{\,}|@{\,\,}r|p{15pt}|p{8pt}|p{15pt}|p{10pt}|p{12pt}|p{17pt}|r@{\,\,}|} \hline
name    & \#latch  & \#pro    & \multicolumn{4}{c|}{Joint verification} & \multicolumn{3}{c|}{\ja-verification by \db{}} \\ \cline{4-7}
        &          & pert.    & \multicolumn{2}{c|}{\Abc} &\multicolumn{2}{c|}{\db}
        & \multicolumn{3}{c|}{with clause re-use} \\ \cline{4-10}
        &          &          & \#false      &        &\#false  &      & time       & \#false   & total \\ 
        &          &          & (\#true)     &  time  &(\#true) & time & limit      & (\#true)  & time \\ \hline
6s104   & 84,925   &  124     &    1\,(0)    & 10\,h  & 1\,(0)   & mem  &  0.3\,h   &  1\,(123) & 2.5\,h \\ \hline
6s260   &  2,179   &   35     &    1\,(0)    & 10\,h  & 1\,(0)   & 10\,h&  0.5\,h    &  1\,(34) &1,686\,s \\ \hline
6s258   &  1,790   &   80     &    25\,(0)   & 10\,h  & 30\,(0)  & 10\,h&  0.3\,h    &  1\,(72)  & 2.4\,h  \\ \hline
6s175   &  7,415   &   3      &    2\,(0)    & 10\,h  & 2\,(0)   &10\,h &  0.3\,h    &  2\,(1)   & 554\,s  \\ \hline
6s207   &  3,012   &   33     &    6\,(0)    & 10\,h  & 10\,(0)  &10\,h &  0.3\,h    &  2\,(31)  & 22\,s \\ \hline   
6s254   &  762     &   14     &    13\,(1)   & 25\,s  & 13\,(1)  &225\,s&  0.3\,h    &  1\,(13)  & 2\,s\\ \hline
6s335   &  1,658   &  61      &    26\,(35)  & 2\,h   & 26\,(35) & 260\,s& 0.3\,h    &  20\,(41) & 56\,s  \\ \hline
6s380   &  5,606   &  897     &  399\,(0)    & 10\,h  & 395\,(0) &10\,h &  0.3\,h    &  3\,(894) & 550\,s \\ \hline
\end{tabular}                
\end{center}
\vspace{-15pt}
\label{tbl:some_failed}
\end{table}


%
%
\subsection{Designs with failing properties}
\label{ssec:fail}
In this section, we describe an experiment where we verified designs
with failing properties. Our objective was to show that solving
properties locally can be much more efficient than globally. The
results are given in Table~\ref{tbl:some_failed}. The first column
provides the name of the benchmark.  The second and third columns give
the number of latches and properties, respectively.  The next two
pairs of columns provide the results of joint verification performed
by \Abc and \db. The first column of the pair gives the number of
false and true properties that \Abc or \db managed to solve within the
time limit. The second column of the pair reports the amount of time
taken by \Abc or \db.  The last three columns report data about
\ja-verification: the time limit per property, the number of false and
true properties solved within the time limit, and the total time taken
by \db. In all tables of experimental sections, the run times that do
not exceed one hour are given in seconds.

For all examples but \ti{6s258}, \ja-verification solved all
properties \ti{locally}.  On~the other hand, for many examples, in
joint verification, only a small fraction of properties were solved by
\db and \Abc \ti{globally}.  Let us consider example \ti{6s207} in
more detail.  \ja-verification solved all properties of \ti{6s207}
fairly quickly generating the debugging set of two properties.  On the
other hand, joint verification by \db proved that ten properties
failed globally within 10~hours.  Since \ja-verification showed that
only two properties failed locally, eight out of those ten failed
properties were true locally. Let $P_i$ be one of those eight
properties.  The \cex found for $P_i$ by joint verification first
falsifies a property of the debugging set. Thus, we do not know if
\ti{there is} a \cex where $P_i$ fails before other
properties. \ja-verification does not determine whether $P_i$ fails
but guarantees that \ti{every} \cex for $P_i$ (if any) first fails
some other property.

%
%
\begin{table}
\small
\caption[caption]{\ti{All properties are true}}
\vspace{-10pt}
\scriptsize
\begin{center}
\begin{tabular}{|p{26pt}|r|r|r|r|r|c|r|} \hline
  name   & \#latch  &  \mbox{\#pro-} &\multicolumn{2}{c|}{Joint verification} &
                                                                               \multicolumn{3}{c|}{\ja-verification by \db}  \\ 
         &          &  pert.          &\multicolumn{2}{c|}{} & \multicolumn{3}{c|}{with clause re-use}   \\\cline{4-8}
         &          &        & \Abc         &  \mbox{\db}                        & time  &\#un-    & total   \\
         &          &        & time     &     time       & limit & solved  &    time  \\ \hline
6s124    & 6,748    &  630   & $>$10\,h &    2.9\,h                 &  0.8\,h & 0 & \tb{1.9\,h}  \\ \hline
6s135    & 2,307    &  340   &  123\,s  &     \tb{335\,s}           &  0.8\,h  & 0 & 746\,s \\ \hline
6s139    & 16,230   &  120   &  4.7\,h  &   \tb{1.7\,h}             &  2.8\,h  & 2 & 6.5~h \\ \hline
6s256    &  3,141   &   5    & $>$10\,h &    \tb{602\,s}            &  2.8\,h  & 1 & 2.9~h \\ \hline
bob12m09 & 285   &   85   & 1,692\,s   &  930\,s                   &  0.8\,h & 0 & \tb{784\,s}  \\ \hline
6s407    & 11,379&   371  & 1.3\,h   &    3.4\,h                 &  0.8\,h  & 0 & \tb{2,077\,s}  \\ \hline
6s273    & 15,544&  42    & 1.8\,s   &   325\,s                  &  0.8\,h  & 0 & \tb{290\,s} \\ \hline
6s275    & 3,196 &  673   & 334\,s   &   \tb{1,154\,s}           &  0.8\,h  & 0 & 1,611\,s \\ \hline

\end{tabular}                
\end{center}
\vspace{-15pt}
\label{tbl:all_true}
\end{table}

%
%
\subsection{Designs where all properties hold}
\label{ssec:hold}

In this subsection, we describe an experiment with eight designs where
all properties were true. The results are given in
Table~\ref{tbl:all_true}. The first three columns are the same as in
Table~\ref{tbl:some_failed}.  The next two columns give run times of
\Abc and \db in joint verification. The last three columns provide
information about \ja-verification: time limit per property, number of
unsolved properties and total run time. The best of the run times
obtained in joint verification and \ja-verification based on \db is
given in bold. In three cases, joint verification based on \Abc was
the fastest but we needed a comparison that uses a uniform setup.

Table~\ref{tbl:all_true} shows that joint verification performed
slightly better. In particular, for benchmarks \ti{6s139} and
\ti{6s256}, \ja-verification failed to solve some properties with the
time limit of 2.8 hours. However, when we verified properties in an
order different from the one of design description, both benchmarks
were solved in time comparable with joint verification.


\section{Studying  \ja-verification In More Detail}
\label{sec:exper_with_sep}

\subsection{Comparison of local and global proofs}
\label{ssec:loc_glob}
In this subsection, we describe an experiment where we compared
separate verification with global and local proofs. Both versions of
separate verification employed clause re-using. (Thus, separate
verification with local proofs is \ja-verification). In
Table~\ref{tbl:fls_lo_gl}, we compare global and local proofs on
benchmarks with failing properties from
Table~\ref{tbl:some_failed}. The first two columns provide the name of
a benchmark and its number of properties. The next two columns specify
the performance of separate verification with global proofs. The first
column of the two shows how many properties were solved within a
10-hour time limit for the entire benchmark. The time limit per
property was the same as in Table~\ref{tbl:some_failed}. The second
column gives the overall time for a benchmark. The last two columns
provide the same information for separate verification with local
proofs. Table~\ref{tbl:fls_lo_gl} shows that separate verification
with local proofs dramatically outperforms the one with global proofs.

%
%
\begin{table}
\small
\caption[caption]{\ti{Separate verification  with global and local proofs
 for examples of Table~\ref{tbl:some_failed}. Time limit per  property is the
 same as in Table~\ref{tbl:some_failed}. The total time limit per benchmark
 is 10 hours}}
\vspace{-10pt}
\scriptsize
\begin{center}
\begin{tabular}{|l|r|c|r|c|r|} \hline
name      &\!\!\#properties\!\!& \multicolumn{2}{c|}{global proofs} & \multicolumn{2}{c|}{local proofs}\\ \cline{3-6}
&            &\mbox{\#unsolved}&     time     &  \mbox{\#unsolved}      & time    \\ \hline
 6s104     & 124   &   123  &     10\,h       &   \tb{0}  &   \tb{2.5\,h}    \\ \hline
 6s260     &  35   &   36   &     10\,h       &   \tb{0}  &   \tb{1,686\,s}  \\ \hline
 6s258     &  80   &   70   &     10\,h       &   \tb{7}  &   \tb{2.4\,h}      \\ \hline
 6s175     &   3   &   1    &     1,070\,s    &   \tb{0}  &   \tb{554\,s}    \\ \hline
 6s207     &   33  &   23   &     7.0\,h      &   \tb{0}  &   \tb{22\,s}     \\ \hline
 6s254     &  14   &   0    &     237\,s      &    0      &   \tb{2\,s}       \\ \hline
 6s335     &   61  &   3    &     3,243\,s    &   \tb{0}  &   \tb{56\,s}        \\ \hline
 6s380     &   897 &  698   &      10\,h      &   \tb{0}  &   \tb{550\,s}    \\ \hline


\end{tabular}
\end{center}
\vspace{-5pt}
\label{tbl:fls_lo_gl}
\end{table}

In Table~\ref{tbl:true_lo_gl}, we compare global and local proofs on
benchmarks from Table~\ref{tbl:all_true} where all properties are
true. The structure of Table~\ref{tbl:true_lo_gl} is the same as that
of Table~\ref{tbl:fls_lo_gl}. Table~\ref{tbl:true_lo_gl} demonstrates
that both versions of separate verification show comparable
performance. A noticeable difference is observed only on benchmarks
6s256 and 6s407. So one can conclude that it is more likely to see the
effect of using local proofs on benchmarks with failed properties.  On
the other hand, the advantage of using local proofs may become more
pronounced even on correct designs if the number of properties is
large (see Section~\ref{sec:parallel}).

%
%
\begin{table}[h]
\small
\caption[caption]{\ti{Separate verification  with global and local proofs
 for examples of Table~\ref{tbl:all_true}. Time limit per  property is the
    same as in Table~\ref{tbl:all_true}}}
\vspace{-10pt}
\scriptsize
\begin{center}
\begin{tabular}{|l|r|c|r|c|r|} \hline
name      &\!\!\#properties\!\!& \multicolumn{2}{c|}{global proofs} & \multicolumn{2}{c|}{local proofs}\\ \cline{3-6}
&            &\mbox{\#unsolved}&     time     &  \mbox{\#unsolved}      & time    \\ \hline
 6s124     & 630   &    0   &    2.1\,h       &  0   &  \tb{1.9\,h}     \\ \hline
 6s135     & 340   &    0   &    764\,s       &  0   &  \tb{746\,s}     \\ \hline
 6s139     & 120   &    2   &    8.1\,h       &  2   &  \tb{6.5\,h}     \\ \hline
 6s256     &  5    &    2   &    5.7\,h       &  \tb{1}   &  \tb{2.9\,h}     \\ \hline
 bob12m09  & 85    &    0   &    809\,s       &  0   &  \tb{784\,s}      \\ \hline
 6s407     & 317   &    5   &    5.4\,h       &  \tb{0}   &  \tb{2,077\,s}      \\ \hline
 6s273     & 42    &    0   &    \tb{278\,s}       &  0   &  290\,s     \\ \hline
 6s275     & 673   &    0   &    \tb{1.542\,s}     &  0   &  1,611\,s     \\ \hline


\end{tabular}
\end{center}
\vspace{-5pt}
\label{tbl:true_lo_gl}
\end{table}

\subsection{Benefit of clause re-using}

%
%
\begin{table}
\small
\caption[caption]{\ti{Re-using strengthening clauses in
    \ja-verification. Time limit per  property is the
    same as in Table~\ref{tbl:all_true}.
    Verification of 6s124, 6s139, and 6s407  (without clause re-use)
    was aborted after 10 hours}}
\vspace{-10pt}
\scriptsize
\begin{center}
\begin{tabular}{|l|r|c|r|c|r|} \hline
name      &\!\!\#properties\!\!& \multicolumn{2}{c|}{without clause re-use} & \multicolumn{2}{c|}{with clause re-use}\\ \cline{3-6}
&            &\mbox{\#unsolved}&     time          &  \mbox{\#unsolved}      & time    \\ \hline
 6s124     & 630   &  505   &       10\,h     & 0    & \tb{1.9\,h} \\ \hline
 6s135     & 340   &  0     &       2.7\,h    & 0    & \tb{746\,s} \\ \hline
 6s139     & 120   &  116   &       10\,h     & 2    & \tb{6.5\,h} \\ \hline
 6s256     &  5    &  0     &     \tb{892\,s} & 1    &     2.9\,h  \\ \hline
 bob12m09  & 85    &  0     &       1.1\,h    & 0    & \tb{784\,s}  \\ \hline
 6s407     & 317   &  270   &       10\,h     & 0    & \tb{2,077\,s}  \\ \hline
 6s273     & 42    &  0     &       1,445\,s  & 0    & \tb{290\,s} \\ \hline
 6s275     & 673   &  0     &       3,273\,s  & 0    & \tb{1,611\,s} \\ \hline


\end{tabular}
\end{center}
\vspace{-15pt}
\label{tbl:reusing}
\end{table}

To illustrate the benefit of re-using strengthening clauses,
Table~\ref{tbl:reusing} compares \ja-verification with and without
re-using strengthening clauses on the examples of
Table~\ref{tbl:all_true}. Table~\ref{tbl:reusing} shows that
\ja-verification with re-using strengthening clauses significantly
outperforms its counterpart. The only exception is \ti{6s256} which
has only five properties to check.

%
%
\begin{table}[h]
\small
\caption[caption]{\ti{\ja-verification with lifting respecting or ignoring property constraints for examples of
    Table~\ref{tbl:some_failed}. Time limit per property is the same
    as in Table~\ref{tbl:some_failed}}}
\vspace{-10pt}
\scriptsize
\begin{center}
\begin{tabular}{|l|r|c|r|c|r|} \hline
name      &\!\!\#properties\!\!& \multicolumn{2}{c|}{respecting prop. constr.} & \multicolumn{2}{c|}{ignoring prop. constr.}\\ \cline{3-6}
&            &\mbox{\#unsolved}&     time     &  \mbox{\#unsolved}      & time    \\ \hline
 6s104     & 124   &    0   &    2.5\,h       &   0       &    2.5\,h        \\ \hline
 6s260     &  35   &    2   &    1\,h         &   \tb{0}  &   \tb{1,686\,s}  \\ \hline
 6s258     &  80   & \tb{0} &  \tb{69\,s}     &   7       &   2.4\,h         \\ \hline
 6s175     &   3   &    0   &  \tb{294\,s}    &   0       &   554\,s         \\ \hline
 6s207     &   33  &    0   &   33\,s         &   0       &   \tb{22\,s}     \\ \hline
 6s254     &  14   &    0   &   2\,s          &   0       &   2\,s            \\ \hline
 6s335     &   61  &    0   &   120\,s        &   0       &   \tb{56\,s}        \\ \hline
 6s380     &   897 &    0   &   878\,s        &   0       &   \tb{550\,s}    \\ \hline


\end{tabular}
\end{center}
\vspace{-5pt}
\label{tbl:fls_lift}
\end{table}

%
%
\subsection{\ja-verification and state-lifting}
\label{ssec:lift}
As we said in Subsection~\ref{ssec:prop_constr}, \db proves property
$P_i$ locally by treating all the properties $P_j$, $j \neq i$ as
constraints. We also mentioned that this may affect the state lifting
procedure used by \pdr. In this subsection, we study this issue
experimentally by considering two versions of \db. In the first
version, property constraints are respected when lifting a state
\pnt{q} whereas in the second version these constraints are
ignored. Respecting and ignoring property constraints means the
following. Let \cof{C}{q} be the cube obtained by lifting state
\pnt{q}. When proving property $P_i$, the first version of \db
guarantees that the states of \cof{C}{q} satisfy $P_j$, $j \neq i$
whereas this is, in general, not true for the second version.

%
%
\begin{table}
\small
\caption[caption]{\ti{\ja-verification with lifting respecting or ignoring property constraints
 for examples of Table~\ref{tbl:all_true}. Time limit per  property is the
    same as in Table~\ref{tbl:all_true}. The total time limit per benchmark is 10 hours}}
\vspace{-10pt}
\scriptsize
\begin{center}
\begin{tabular}{|l|r|c|r|c|r|} \hline
name      &\!\!\#properties\!\!& \multicolumn{2}{c|}{respecting prop. constr.} & \multicolumn{2}{c|}{ignoring prop. constr.}\\ \cline{3-6}
&            &\mbox{\#unsolved}&     time     &  \mbox{\#unsolved}      & time    \\ \hline
 6s124     & 630   &  618   &       10\,h     & \tb{0}  &  \tb{1.9\,h}     \\ \hline
 6s135     & 340   &  248   &       10\,h     & \tb{0}  &  \tb{746\,s}     \\ \hline
 6s139     & 120   &  98    &       10\,h     & \tb{2}  &  \tb{6.5\,h}     \\ \hline
 6s256     &  5    & \tb{0} & \tb{1,282\,s}   &  1      &  2.9\,h          \\ \hline
 bob12m09  & 85    &  0     &  1,070\,s       &  0      &  \tb{784\,s}     \\ \hline
 6s407     & 317   &  0     &  1.0\,h         &  0      &  \tb{2,077\,s}   \\ \hline
 6s273     & 42    &  0     &  537\,s         &  0      &  \tb{290\,s}   \\ \hline
 6s275     & 673   &  0     &  6.4\,h         &  0      &  \tb{1,611\,s}     \\ \hline


\end{tabular}
\end{center}
\vspace{-5pt}
\label{tbl:true_lift}
\end{table}

Table~\ref{tbl:fls_lift} shows the results of both versions of \db on
the failed designs of Table~\ref{tbl:some_failed}. The first two
columns give the name of a benchmark and the total number of
properties. The next two columns show the results of the first version
of \db. These columns give the number of unsolved properties and the
total run time. The last two columns provide the same information for
the second version. The results of Table~\ref{tbl:fls_lift} show that
both versions of \db have comparable performance.

Table~\ref{tbl:true_lift} shows the results of both versions of \db on
the designs of Table~\ref{tbl:all_true} where all properties are true.
The structure of this table is the same as that of
Table~\ref{tbl:fls_lift}. Table~\ref{tbl:true_lift} shows that the
first version is faster than the second version only on one
benchmark. On the other benchmarks the second version outperforms the
first version, sometimes quite dramatically.

\section{\ja-Verification And Parallel Computing}
\label{sec:parallel}
Intuitively, \ja-verification can significantly benefit from parallel
computing. In this section, we give the results of a simple experiment
substantiating this intuition. Our interest in a discussion of
\ja-verification in the context of parallel computing is based on the
following two observations.  Let \mb{P} denote the set of properties
\s{P_1,\dots,P_k} to prove. The first observation is that the larger
$\mb{P}$, the easier proving $P_i\in \mb{P}$ locally due to growing
number of constraints $P_j$, $j \neq i$. The second observation is
that the larger \mb{P}, the smaller an inductive invariant for
property $P_i \in \mb{P}$. Hence, when proving different properties of
\mb{P} locally, the exchange of information (in the form of
strengthening clauses) reduces as \mb{P} grows. The observations above
suggest that by proving properties in parallel one can significantly
decrease verification time.

%
%
\begin{table}
\small
\caption[caption]{\ti{Verification of single properties of benchmark 6s289 (10,789 properties)
    using global and local proofs}}
\vspace{-10pt}
\scriptsize
\begin{center}
\begin{tabular}{|r|c|r|c|r|} \hline
  prop. \!\!& \multicolumn{2}{c|}{global proof} & \multicolumn{2}{c|}{local proof}\\ \cline{2-5}
  index     &\mbox{\#time frames}&     time     &  \mbox{\#time frames}      & time    \\ \hline
  20          &   11   &    213\,s   &  \tb{1}   &   \tb{4.0\,s}        \\ \hline
  137         &   13   &    289\,s   &   \tb{1}   &  \tb{3.0\,s}        \\ \hline
  500         &   12   &    370\,s   &   \tb{1}   &  \tb{7.7\,s}        \\ \hline
  1,001       &   10  &     10\,s    &   \tb{1}  &   \tb{2.5\,s}       \\ \hline
  1,310       &   9    &    30\,s    &   \tb{1}   &  \tb{2.4\,s}        \\ \hline
  2,678       &   9    &    21\,s    &   \tb{1}   &  \tb{2.3\,s}        \\ \hline
  4,789       &   14   &    32\,s    &   \tb{1}   &  \tb{3.5\,s}        \\ \hline
  6,600       &   14   &    75\,s    &   \tb{1}   &  \tb{2.8\,s}          \\ \hline
  10,002      &   10   &    9.9\,s   &   \tb{1}   &  \tb{2.7\,s}          \\ \hline
  \ti{max}    &   14   &    370\,s   &   \tb{1}   &  \tb{7.7\,s}           \\ \hline
\end{tabular}
\end{center}
\vspace{-5pt}
\label{tbl:sep_props}
\end{table}

In Table~\ref{tbl:sep_props}, we report the results of a simple
experiment where we randomly picked individual properties of benchmark
6s289 and proved them both locally and globally.  (The total number of
properties of 6s289 is 10,789.)  The proofs were generated
\ti{independently} of each other i.e. there was no exchange of
strengthening clauses.
The index of selected property is shown in the first column of
Table~\ref{tbl:sep_props}. The next two columns provide information
about the performance of \db when proving the selected property
globally. The first column gives the number of time frames \db had to
unfold. The second column of the two shows the run time taken by
\db. The next pair of columns provide the same information when \db
proved the selected property locally. Table~\ref{tbl:sep_props} shows
that proving the properties we tried locally was very easy.  Assume
that finding a local proof for each of the remaining properties of
6s289 takes a very small amount of time as well. Then, if one had,
say, 10,789 processors to prove each property $P_i$ on a separate
processor, verification would be finished in a matter of seconds.

\section{Related Work}
\label{sec:bck_grnd}

We found only a few references to research on multi-property
verification. In~\cite{abc_multi}, some modifications of \Abc{} are
presented that let it handle multi-property designs.
In~\cite{grouping, mprop_date11}, the idea of grouping similar
properties and solving them together is introduced. The similarity of
properties is decided based on design structure (e.g., properties with
similar cones of influence are considered similar). The main
difference of this approach from ours is that the latter is purely
\ti{semantic}.  Thus, the optimizations of separate verification we
consider (local proofs and re-using strengthening clauses) can be
incorporated in any structure-aware approach.  One further difference is
that the idea of grouping favors \ti{correct}
designs. Grouping may not work well for designs with broken properties
that fail for different reasons and thus have vastly different \cex{}s.

Assume-guarantee reasoning is an important method for compositional
verification~\cite{ag_jones,ag_pnueli}. It reduces verification of the
whole system to checking properties of its components under some
assumptions. To guarantee the correctness of verification one needs to
prove these assumptions true.  As we mentioned earlier, 
\ja-verification uses yet-unproven properties as assumptions
without subsequent justification.  This is
achieved by our particular formulation of multi-property
verification. Instead of proving or refuting every property, 
\ja-verification builds a subset of failed properties that are the
first to break or proves that this subset is empty.


\section{Conclusions}
\label{sec:concl}

We consider the problem of verifying multiple properties $P_1,\ldots,P_k$
of the same design. We make a case for separate verification where
properties are proved one by one as opposed to joint verification where
the aggregate property $P_1 \wedge \ldots \wedge P_k$ is used. Our
approach is purely semantic, i.e., we do not rely on
any structural features a design may or may not have.

We introduce a novel variant of separate verification called
\ja-verification. \ja-verification checks if $P_i$ holds locally, i.e.,
under the assumption that all other properties are true. We show that if
all properties hold locally, they also hold globally, i.e., without any
assumptions. Instead of finding the set of all failed properties,
\ja-verification identifies a ``debugging'' subset. The properties in
the debugging subset highlight design behaviors that need to be fixed
first, which can yield substantial time savings in the
design-verification cycle.

We experimentally compare conventional joint verification and
\ja-verification.  We~give examples of designs with failed properties
where \ja-verification dramatically outperforms its counterpart,
especially for designs where a small debugging set $D$ exists. For
these designs, one needs to find only $|D|$ \cex{}s which are
typically shallow.  Computation of deeper \cex{}s for false properties
that are not in $D$ is replaced with proving them true locally.
Re-using inductive invariants generated for individual properties that
are locally true significantly speeds up \ja-verification. In
particular, for correct designs, it makes \ja-verification competitive
with joint verification even for benchmarks that favor the latter.


\bibliographystyle{IEEEtran}
\bibliography{short_sat,l1ocal}


\end{document}